\documentclass[%
 aip,
 jmp,%
 amsmath,amssymb,
preprint,%
]{revtex4-1}

\usepackage{graphicx}
\usepackage{dcolumn}
\usepackage{bm}

\def\jgr{{J.~Geophys.~Res.}}%

\def\grl{{Geophys.~Res.~Lett.}}%
\def\mnras{{MNRAS}}%
\def\apj{{Astrophys. J.}}%
\def\apjl{{Astrophys. J.}}%

\newcommand {\vect}[1]{\mbox{\boldmath $#1$}}

\newcommand {\pdif}[3][]{\frac{\partial^{#1}#2}{\partial#3^{#1}}}

\makeatletter
\def\mart{\@ifnextchar[{\mart@@}{\mart@}}
\def\mart@@[#1]#2{\sqrt[#1]{\mathstrut{#2}}}
\def\mart@#1{\sqrt{\mathstrut{#1}}}
\makeatother
\newcommand {\Alfven}{Alfv\'{e}n}

\begin{document}


\title{Boosting Magnetic Reconnection by Viscosity and Thermal Conduction}

\author{Takashi Minoshima}
\email{minoshim@jamstec.go.jp}
\affiliation{Department of Mathematical Science and Advanced Technology, Japan Agency for Marine-Earth Science and Technology, 3173-25, Syowa-machi, Kanazawaku, Yokohama 236-0001, Japan}
\author{Takahiro Miyoshi}
\affiliation{Graduate School of Science, Hiroshima University, 1-3-1 Kagamiyama, Higashi-hiroshima 739-8526, Japan}
\author{Shinsuke Imada}
\affiliation{Institute for Space-Earth Environmental Research, Nagoya University, Furo-cho, Chikusa-ku, Nagoya 464-8601, Japan}
\date{\today}

\begin{abstract}
Nonlinear evolution of magnetic reconnection is investigated by means of magnetohydrodynamic simulations including uniform resistivity, uniform viscosity, and anisotropic thermal conduction.
When viscosity exceeds resistivity (the magnetic Prandtl number $Pr_m > 1$), the viscous dissipation dominates outflow dynamics and leads to the decrease in the plasma density inside a current sheet.
The low-density current sheet supports the excitation of the vortex.
The thickness of the vortex is broader than that of the current for $Pr_m > 1$.
The broader vortex flow more efficiently carries the upstream magnetic flux toward the reconnection region, and consequently boosts the reconnection.
The reconnection rate increases with viscosity provided that thermal conduction is fast enough to take away the thermal energy increased by the viscous dissipation (the fluid Prandtl number $Pr < 1$).
The result suggests the need to control the Prandtl numbers for the reconnection against the conventional resistive model.
\end{abstract}

\maketitle


\section{INTRODUCTION}\label{sec:introduction}
Magnetic reconnection is one of the most fundamental processes in plasma physics, in which 
stored magnetic energy is rapidly released and converted into kinetic and internal energies through the change of magnetic field topology.
It is widely believed to play a major role in explosive phenomena such as magnetospheric substorms and stellar flares.
The reconnection intrinsically contains a hierarchical structure ranging from the fully kinetic scale to the magnetohydrodynamic (MHD) scale.
In order to identify the essential physics necessary to model the reconnection, \citealt{2001JGR...106.3715B} conducted numerical simulations with a variety of codes from kinetic codes to conventional resistive MHD codes.
They showed that only the MHD simulation with uniform resistivity fails to trigger fast reconnection, indicating that resistive MHD would be insufficient to model it.

The role of resistive dissipation on the reconnection has been extensively investigated in the framework of MHD.
A classical Sweet-Parker model predicts the rate of the reconnection proportional to the square root of resistivity that is too slow to account for observed phenomena.
Subsequent studies have demonstrated that the Sweet-Parker type current sheet undergoes secondary instabilities for sufficiently small resistivity (large Lundquist number) \citep{1986PhFl...29.1520B,2005PhRvL..95w5003L,2007PhPl...14j0703L,2009PhRvL.103j5004S}.
The resulting reconnection rate seems to be independent of resistivity \citep{2008PhRvL.100w5001L,2009PhPl...16k2102B,2012PhPl...19d2303L,2015PhPl...22j0706S}. 
Meanwhile, the impact of other dissipation processes should be discussed.
We focus on viscosity and heat transfer.


Viscosity might support the nonlinear evolution of the reconnection \citep{2009PhPl...16f0701B} whereas it suppresses the linear growth \citep{1987PhFl...30.1734P}.
The ratio of kinematic viscosity to resistivity is defined as the magnetic Prandtl number, which relates the dissipation scale of vortex to current.
Resistive MHD assumes this number to be zero, meaning that the vortex scale is negligible small compared with the current scale.
However, it is not necessarily true in actual plasma environments \citep{2015ApJ...801..145T};
the number can be much larger than unity in a classical Spitzer model for hot tenuous plasmas \citep{1962pfig.book.....S}.
Numerical simulations have demonstrated that it affects the nonlinear evolution of MHD phenomena such as small-scale turbulence and dynamo \citep{2004ApJ...612..276S,2007MNRAS.378.1471L,2014ApJ...791...12B,2015ApJ...808..54M}.
It may also impact on the reconnection in which small-scale dissipation processes eventually result in large-scale evolution.
In the kinetic reconnection composed of collisionless ions and electrons, the reconnection region has a two-scale structure of broad ion diffusion region and narrow electron diffusion region embedded there \citep{1998JGR...103.9165S,2001JGR...106.3721H}.
This structure may be measured as broad vortex and narrow current layers from the viewpoint of MHD, because the momentum and the current are predominantly sustained by ions and electrons, respectively.

Heat transfer is associated with the reconnection.
High-energy particles are produced in the vicinity of a reconnection site and stream along a magnetic field line during the collisionless reconnection \citep{2001JGR...10625979H,2006JGRA..111.9216F,2007JGRA..112.3202I}.
In the solar flare, thermal conduction is effective along a magnetic field line and may affect the evolution of the collisional reconnection \citep{1997ApJ...474L..61Y,1999ApJ...513..516C,2012ApJ...758...20N}.
Including heat transfer increases compressibility that can enhance the reconnection \citep{2011PhPl...18d2104H,2011PhPl...18k1202B,2012PhPl...19h2109B}.
The fluid Prandtl number, the ratio of kinematic viscosity to temperature conductivity, is around $10^{-3}$ in the Spitzer model.
Therefore, actual plasmas can have the following inequality for the timescale of three diffusion processes, $\tau_{\rm heat} < \tau_{\rm viscous} < \tau_{\rm resistive}$.

In order to ascertain the effect of viscosity and heat transfer on the nonlinear evolution of the reconnection, we conduct two-dimensional MHD simulations including viscous dissipation and anisotropic thermal conduction as well as resistive dissipation.

\section{MODEL}\label{sec:numerical-model}
The governing equations are fully-compressible visco-resistive MHD equations coupled with anisotropic thermal conduction, 
\begin{eqnarray}
&&\pdif{\rho}{t} + \nabla \cdot \left(\rho \vect{u}\right) = 0,\label{eq:1}\\
&&\pdif{\left(\rho \vect{u}\right)}{t} + \nabla \cdot \left[ \rho \vect{u}\vect{u} + \left(P+\frac{B^2}{2}\right)\vect{\rm I} - \vect{B}\vect{B} - \rho \nu \vect{\rm S}\right]=0,\label{eq:2}\\
&&\pdif{\vect{B}}{t} + \nabla \times \vect{E} = 0,\label{eq:3}\\
&&\pdif{e}{t} + \nabla \cdot \left[\left(\frac{\rho u^2}{2} + \frac{\gamma P}{\gamma-1}\right)\vect{u} + \vect{E}\times\vect{B}-\rho \nu \vect{\rm S} \cdot \vect{u} - \vect{\rm \kappa} \cdot \nabla \frac{P}{\rho} \right] = 0,\label{eq:4}\\
&&\vect{E} = -\vect{u} \times \vect{B} + \eta \vect{j}, \vect{j} = \nabla \times \vect{B},\label{eq:5}\\
&&P = \left(\gamma-1\right)\left(e-\frac{\rho u^2}{2} - \frac{B^2}{2}\right),\label{eq:6}\\
&&\vect{\rm S} = \sum_{i,j} \sigma_{ij}\vect{e}_i\vect{e}_j,
\sigma_{ii}=\frac{2}{3}\left(2\pdif{u_i}{x_i}-\pdif{u_j}{x_j}\right),\sigma_{ij}=\sigma_{ji}=\left(\pdif{u_i}{x_j}+\pdif{u_j}{x_i}\right),\label{eq:7}\\
&&\vect{\rm \kappa} = \frac{\alpha \rho}{\gamma-1} \sum_{i,j} \frac{B_i B_j}{|\vect{B}|^2+\epsilon}\vect{e}_i\vect{e}_j,\label{eq:8}
\end{eqnarray}
where $\nu$, $\eta$, and $\alpha$ denote the kinematic viscosity, the resistivity, and the temperature conductivity, $\vect{\rm I}$ and $\vect{\rm S}$ the unit and the strain rate tensors, $\vect{\rm \kappa}$ the anisotropic thermal conductivity tensor working only along the magnetic field line,
 $\epsilon = 10^{-6}$ a small value to avoid the division by zero,
 and other symbols have their usual meanings.
{The magnetic permeability is assumed to be unity.}
The three diffusion coefficients $\nu,\eta,\alpha$ are assumed to be uniform and constant.
We use an adiabatic index $\gamma = 5/3$.

The equations are advanced based on an operator splitting technique.
The ideal MHD part is solved by a nonlinear third-order finite difference scheme coupled with the HLLD approximate Riemann solver \citep{2005JCoPh.208..315M}.
The solenoidal condition for magnetic field is guaranteed by the HLLD Upwind Constrained Transport method \citep{2015ApJ...808..54M}.
We use a fourth-order central difference in space and a third-order Runge-Kutta time integration for the resistive dissipation term.
The viscous dissipation and the thermal conduction terms are discretized into second-order in space and time, and are updated by the multi-color Gauss-Seidel method.

The initial condition is a Harris current sheet configuration on background stationary plasma, $\rho=(\rho_0-\delta \rho)\cosh^{-2}(y/\lambda)+\delta \rho,\vect{u}=0,\vect{B}=(B_0 \tanh(y/\lambda),0,0),P=(B_0^2/2)(\rho/\rho_0)$, where $\delta \rho = 0.2 \rho_0$ is the background plasma density and $\lambda$ is the thickness of the current sheet. 
The quantities are normalized so that $\rho_0 = \lambda = B_0 = 1$.
We initiate the reconnection by adding a small perturbation to the flux function $\delta A_z = 0.1\lambda B_0 \exp[-(x^2+y^2)/(2\lambda)^2]$ and a small (1\%) uniform random perturbation to $u_y$ inside the current sheet.
The three diffusion coefficients are expressed as dimensionless numbers of the magnetic and fluid Reynolds numbers, $(R_m,R_e) = (\lambda V_{\rm A0}/\eta,\lambda V_{\rm A0}/\nu)$, and the magnetic and fluid Prandtl numbers, $(Pr_m,Pr)=(\nu/\eta,\nu/\alpha)$, where $V_{\rm A0}=B_0/\mart{\rho_0}$ is the {\Alfven} speed.
We fix a resistivity value of $10^{-3} \; (R_m = 10^{3}) $ and vary the kinematic viscosity and the temperature conductivity.
The rectangular simulation domain $[0,64\lambda]\times[-8\lambda,8\lambda]$ is resolved with uniformly-spaced grid points of $8,192\times2,048$.
The symmetric boundary and the conducting wall boundary are set in $x$ and $y$ directions.
We confirm that doubling the resolution in space and time does not alter our conclusion.

\section{RESULT}\label{sec:result}
\begin{figure}[htbp]
 \includegraphics[clip,angle=0,scale=0.45]{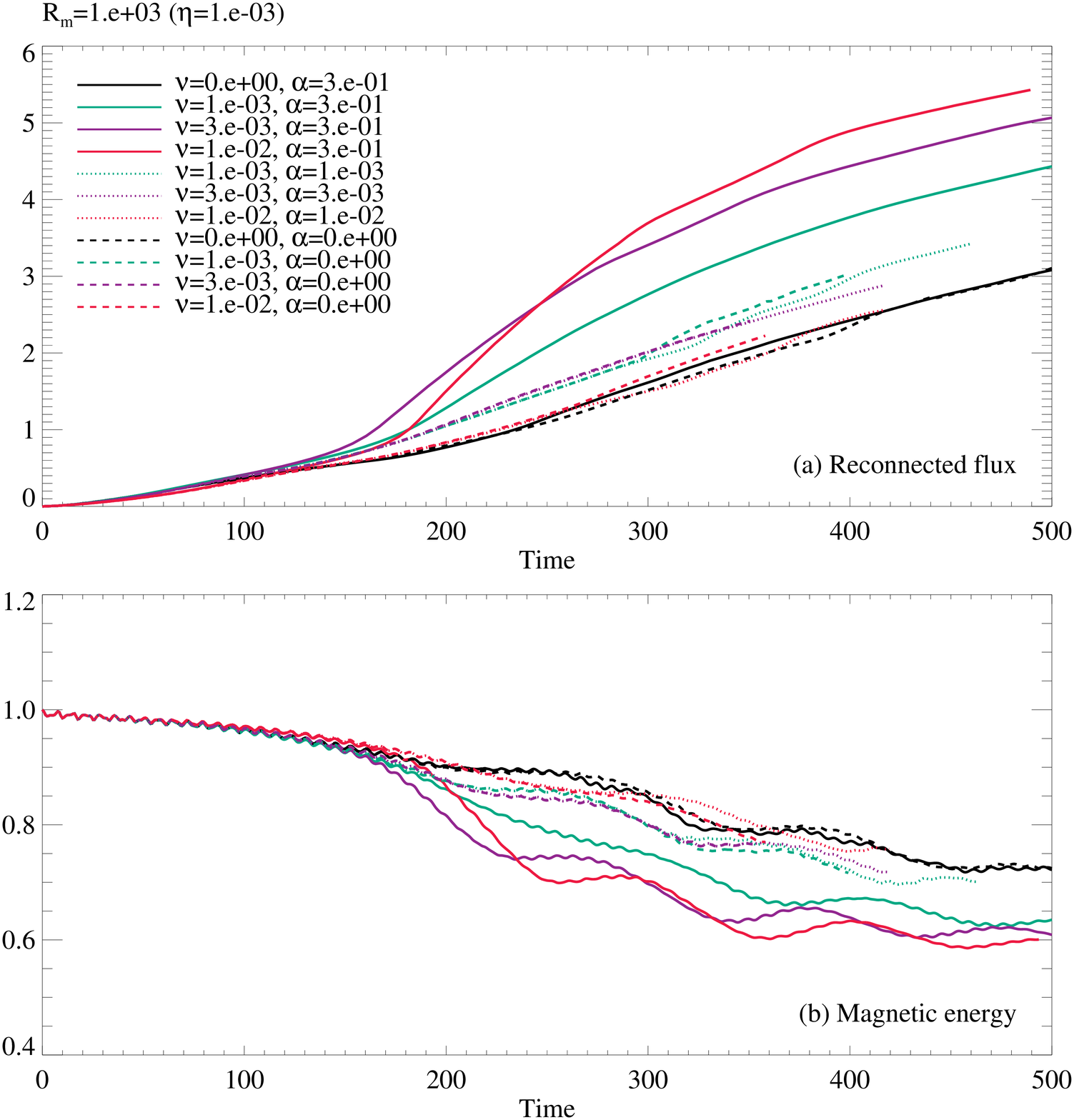}
\caption{Time profile of reconnection; (a) the amount of reconnected flux at the reconnection site, and (b) spatially-integrated magnetic energy. The various colors represent simulation results with different viscosity. The solid, dotted, and dashed lines denote the results including thermal conduction with $Pr < 1$ and $Pr = 1$, and excluding the conduction.}
\label{fig:mflux_btene}
\end{figure}

\begin{figure}[htbp]
 \includegraphics[clip,angle=0,scale=0.45]{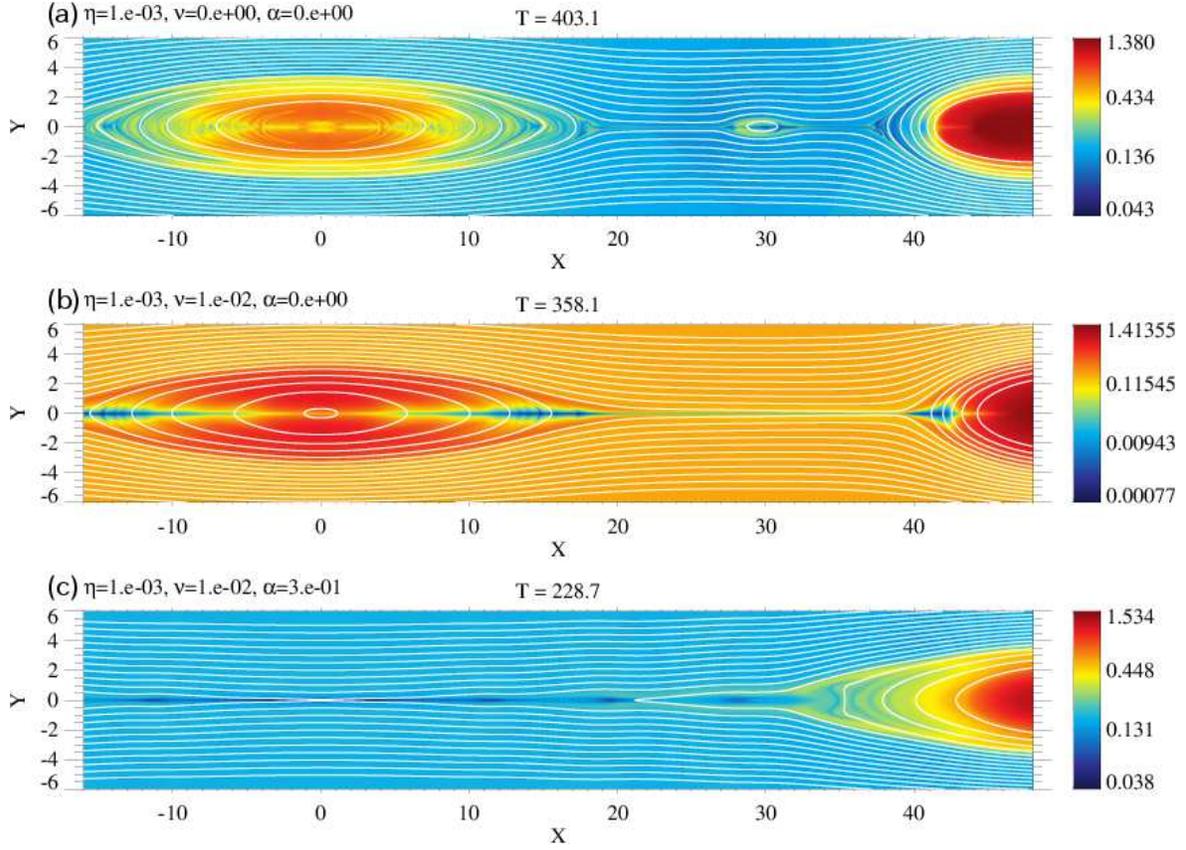}
\caption{Two-dimensional density distribution at which the reconnected flux is about the same for each case; (a) the resistive case, (b) the visco-resistive case $(Pr_m = 10)$, and (c) the visco-resistive case including thermal conduction $(Pr_m = 10, Pr = 1/30)$.}
\label{fig:density_distri}
\end{figure}

We identify the dominant reconnection site by finding the minimum in $x$ of magnetic flux $\int |B_x|dy$, and then measure the amount of the magnetic flux reconnected there.
The time evolution of the reconnected flux is shown in Figure \ref{fig:mflux_btene}(a).
Its slope corresponds to the local reconnection rate.
As a global measure of the reconnection, Figure \ref{fig:mflux_btene}(b) shows the time evolution of magnetic energy integrated over a whole domain.
We use viscosity values of $\nu = 0,10^{-3},3\times 10^{-3},10^{-2} \; (Pr_m=0,1,3,10)$, which are indicated as black, green, magenda, and red lines, respectively.
The temperature conductivity values are set to $\alpha = 0.3 \; (Pr < 1)$, $\alpha = \nu \; (Pr = 1)$, and $\alpha = 0$, which are shown as solid, dotted, and dashed lines, respectively.
Figure \ref{fig:density_distri} shows the plasma density distribution in three different cases, (a) the resistive case $(\nu=\alpha=0)$, (b) the visco-resistive case $(Pr_m=10)$, and (c) the visco-resistive case including the conduction $(Pr_m=10,Pr=1/30)$.

In all cases the current sheet gets thin down after the passage of the initial perturbation.
The length $L$ and the thickness $\delta$ of the current sheet becomes $ 20 \lambda$ and $0.06 \lambda$ at $T=300$ in the resistive case.
Subsequently, an elongated Sweet-Parker type current sheet undergoes secondary tearing instability.
Secondary plasmoids are observed around $x=0$ and $x=30$ in Figure \ref{fig:density_distri}(a).
However, the number of plasmoids is small within the present simulation domain and time, and their impact on the reconnection rate is limited in this case.

The visco-resistive simulations without thermal conduction show that moderate viscosity accelerates the evolution around $T=150$ prior to the onset of the secondary instability in comparison with the resistive case (dashed green and magenda lines in Figure \ref{fig:mflux_btene}).
The acceleration is weakened in the most viscous case (dashed red lines).
A secondary plasmoid is less observed for larger viscosity although the elongated current sheet is formed.
Thermal conduction has little effect on the evolution in the resistive case and the visco-resistive cases as long as $Pr \geq 1$.
Their time profiles are almost same as the cases without the conduction (solid black and dotted lines).


Surprisingly, the reconnection is enhanced when the kinematic viscosity is larger than the resistivity $(Pr_m \geq 1)$ and the temperature conductivity is further larger than the viscosity $(Pr < 1)$.
In the case $(Pr_m,Pr)=(10,1/30)$, the reconnection is initiated slower than the other cases, followed by the explosive onset at $T=180$ (solid red lines in Figure \ref{fig:mflux_btene}).
Figure \ref{fig:density_distri}(c) shows that the spatial structure is considerably different from the previous cases.
A single X-type reconnection site is formed around $x=0$ and the current is localized there.
The outflow speed reaches the {\Alfven} speed measured in the inflow region.
Similar evolution is observed in all three cases for $Pr_m \geq 1$ and $Pr < 1$.
Contrary to the linear theory, the peak reconnection rate increases with the viscosity, $0.005,0.007,0.01,0.013$ for $\nu=0,10^{-3},3\times 10^{-3},10^{-2}$, indicating that viscosity plays a key role in boosting the reconnection.

\begin{figure}[htbp]
 \includegraphics[clip,angle=0,scale=0.7]{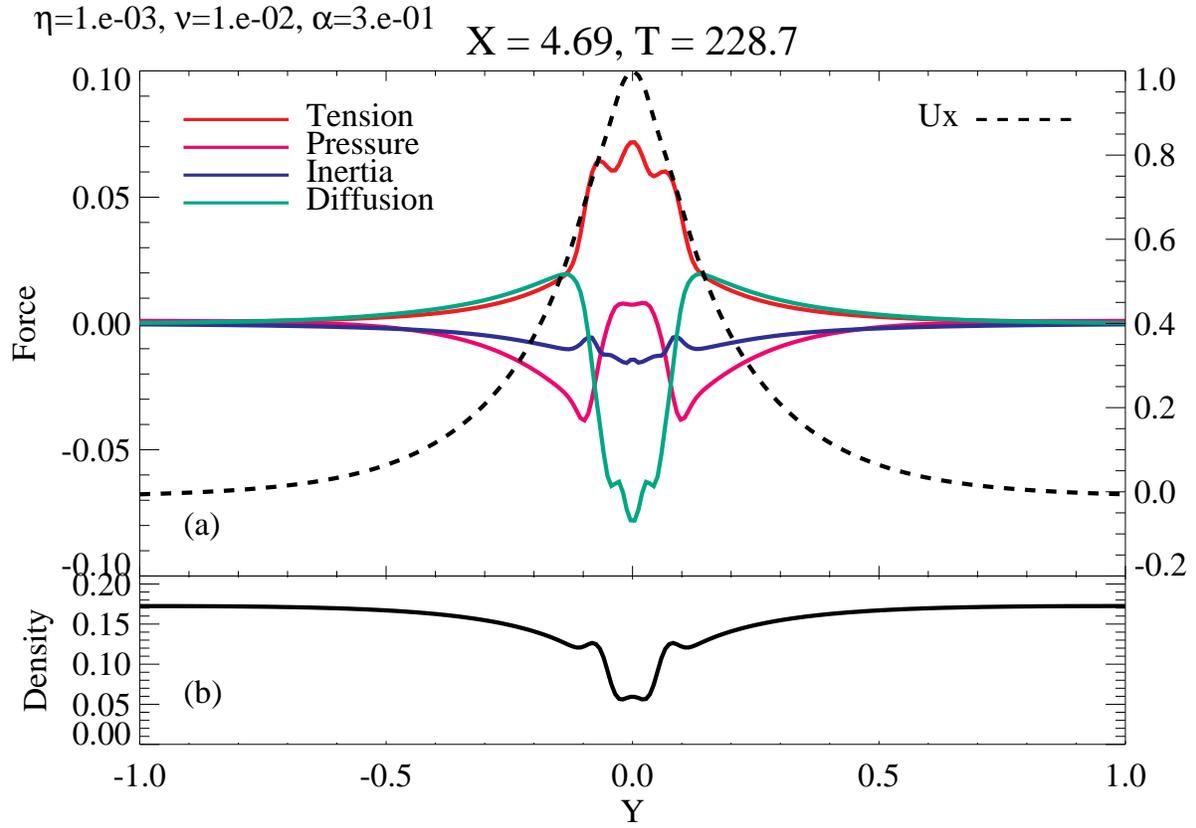}
\caption{One-dimensional distribution across the outflow in the visco-resistive case including thermal conduction $(Pr_m = 10,Pr =1/30)$. (a) The magnetic tension (red), the pressure gradient (pink), the inertia (blue), and the viscous dissipation (green) forces on the outflow momentum. The dashed line represents the outflow profile. (b) The density profile.}
\label{fig:cut1d}
\end{figure}

Figure \ref{fig:cut1d}(a) shows the one-dimensional distribution across the outflow at $x=4.7$ at $T=229$ in the visco-resistive case including the conduction.
The outflow is predominantly accelerated by the magnetic tension force and is decelerated by the viscous dissipation force.
{We then approximate the equilibrium in the current sheet as $\partial(B_x B_y)/\partial y + \nu (\partial/\partial y) \rho \partial u_x / \partial y = 0$ (remaining dissipation terms are relatively small), leading to
\begin{eqnarray}
B_{\rm in} B_{\rm out} = \nu \rho_{\rm out} \frac{u_{\rm out}}{\delta}\label{eq:13}, 
\end{eqnarray}
where quantities with the subscript in(out) are measured in the inflow(outflow) region.
By combining it with the induction and continuity equations,
\begin{eqnarray}
 u_{\rm in} B_{\rm in} = u_{\rm out} B_{\rm out},\label{eq:14}\\
\rho_{\rm in} u_{\rm in} L = \rho_{\rm out} u_{\rm out} \delta,\label{eq:15}
\end{eqnarray}
 the aspect ratio of the reconnection region is dominated by viscosity \citep{1984PhFl...27..137P},
\begin{eqnarray}
L/\delta = \mart{(V_{\rm A,in}L/\nu)(V_{\rm A,in}/u_{\rm out})},\label{eq:16}
\end{eqnarray}
where $V_{\rm A,in}=B_{\rm in}/\mart{\rho_{\rm in}}$ is the {\Alfven} speed in the inflow region.}
In the classical Sweet-Parker model, the outflow is decelerated by the inertia force and the aspect ratio is proportional to the inverse square root of resistivity.

\begin{figure}[htbp]
 \includegraphics[clip,angle=0,scale=0.45]{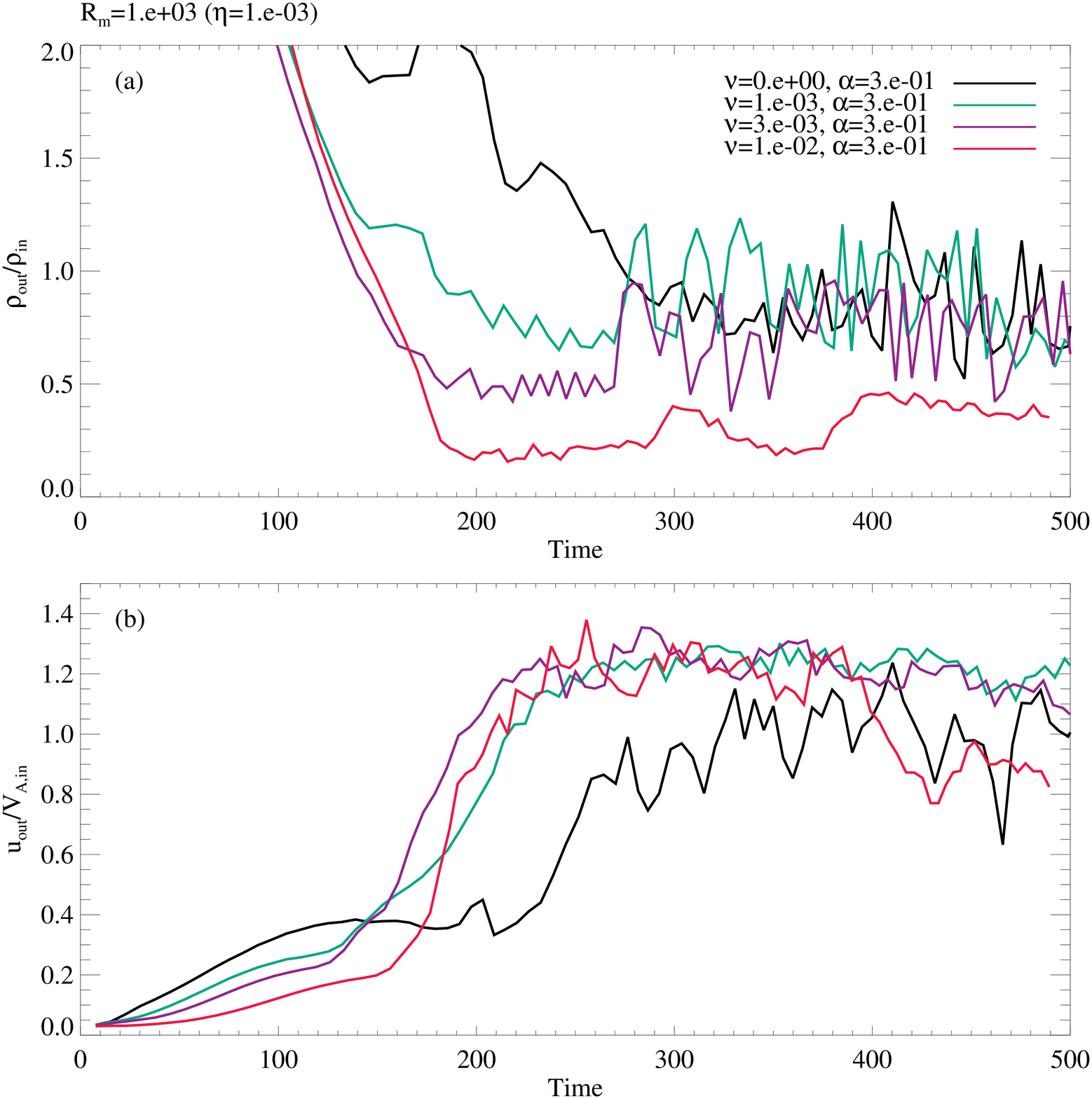}
\caption{Time profile of (a) the minimum outflow density normalized by the inflow density, and (b) the peak outflow velocity normalized by the inflow {\Alfven} velocity. The various colors represent simulation results with different viscosity. The thermal conductivity value $\alpha=0.3$ is adopted.}
\label{fig:ro_ux_out}
\end{figure}

The inflow speed in the visco-resistive case follows $u_{\rm in} = \eta/\delta$ similar to the resistive case.
Consequently, we reduce the continuity equation (\ref{eq:15}) to
\begin{eqnarray}
\rho_{\rm out}/\rho_{\rm in} = Pr_m^{-1} (u_{\rm out}/V_{\rm A,in})^{-2}.\label{eq:17} 
\end{eqnarray}
{The plasma can be diluted for $Pr_m > 1$ if $u_{\rm out} = V_{\rm A,in}$ holds.
Figure \ref{fig:ro_ux_out} supports this relation.
In the nonlinear stage $(T > 200)$, (a) the outflow density decreases with increasing viscosity whereas (b) the outflow velocity $u_{\rm out} \sim V_{\rm A,in}$ is insensitive to it.
The outflow density falls below the inflow density and subsequently the explosive reconnection is triggered.
We also observe the density decreasing from the upstream toward the center of the current sheet in Figure \ref{fig:cut1d}(b).
It rather increases toward the sheet in the resistive case.}

\begin{figure}[htbp]
 \includegraphics[clip,angle=0,scale=0.45]{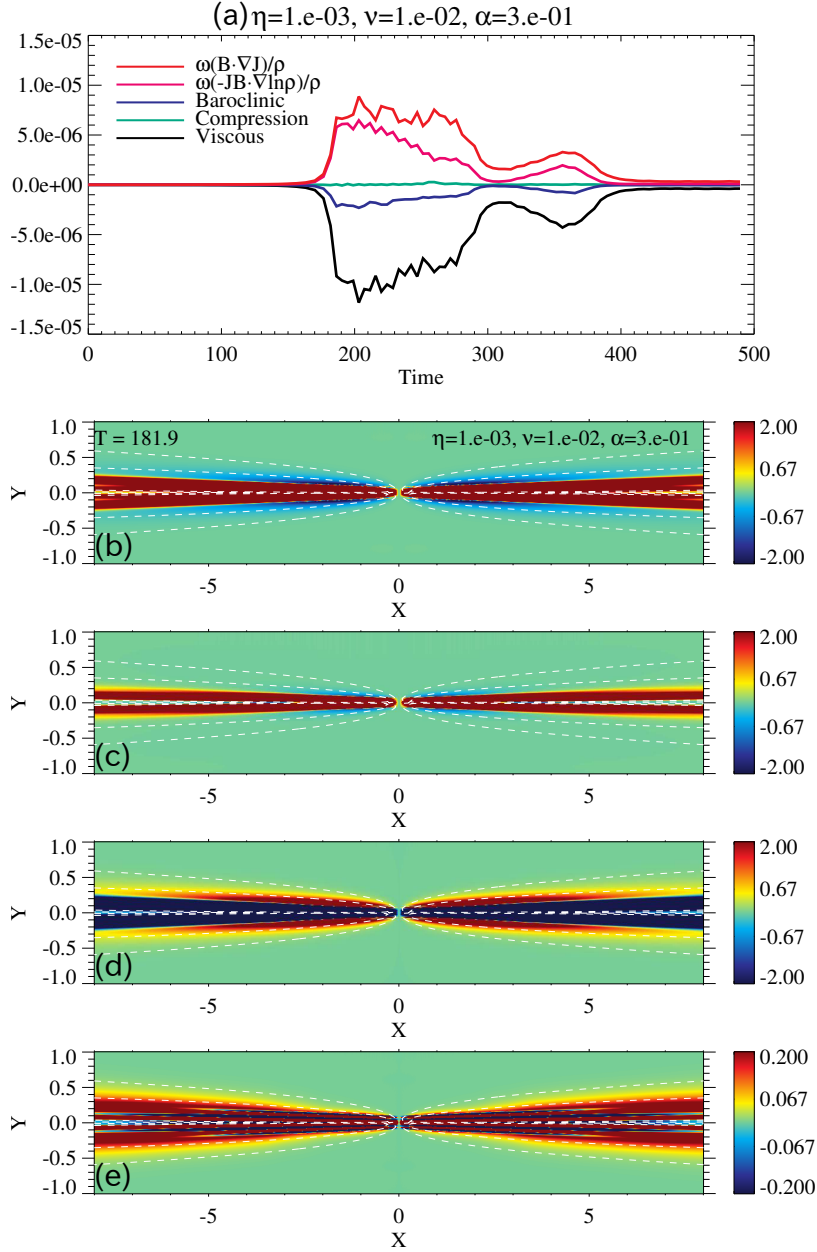}
\caption{Enstrophy equation in the visco-resistive case including thermal conduction $(Pr_m = 10,Pr =1/30)$. (a) Time profile of spatially-integrated terms.  The red and pink lines denote the Lorentz terms related to the gradient of the current and the density. The blue, green, and black lines correspond to the baroclinic, the compression, and the viscous terms. (b-e) Two-dimensional distribution at the onset $(T=182)$; the Lorentz terms related to the gradient of the current and the density, the viscous term, and the whole. The dashed lines denote the enstrophy (10\% and 1\% of the maximum value). The color scale is adjusted so as to visualize diffuse regions.}
\label{fig:voz_prof}
\end{figure}

The essence of the reconnection is the excitation of a quadrupolar vortex around a reconnection site as well as the dissipation of current. 
We investigate the excitation mechanism of the enstrophy $Q = |\vect{\omega}|^2/2$ where $\vect{\omega} = \nabla \times \vect{u}$ is the vorticity.
In the two-dimensional configuration without the out-of-plane magnetic field (that is, $\vect{\omega}=\omega_z \vect{e_z}, \vect{j}=j_z\vect{e_z}$), some vector identities reduce the enstrophy equation to
\begin{eqnarray}
 \pdif{Q}{t} &=& -\left[\nabla \cdot (\vect{u}Q)+Q(\nabla \cdot \vect{u})\right] + \left[\frac{\vect{\omega}}{\rho} \cdot (\vect{B} \cdot \nabla)\vect{j} -\frac{(\vect{\omega}\cdot \vect{j})}{\rho^2}(\vect{B}\cdot \nabla \rho) \right]\nonumber \\
&& -\frac{\vect{\omega}}{\rho^2}\cdot (\nabla P \times \nabla \rho)+\vect{\omega} \cdot \nabla \times \left(\frac{1}{\rho} \nabla \cdot \rho \nu \vect{\rm S}\right).\label{eq:12}
\end{eqnarray}
The equation consists of the inertia (advection and compression) term, the Lorentz term, the baroclinic term, and the viscous term.
The Lorentz term is reduced to two terms; the first proportional to the gradient of the current, and the second proportional to the gradient of the density.
The first Lorentz term is the primary source for the enstrophy under the current sheet configuration since the out-of-plane vector $(\vect{B} \cdot \nabla)\vect{j}$  is parallel to the quadrupolar vortex.
Figure \ref{fig:voz_prof}(a) shows the time profile of the spatially-integrated terms in the visco-resistive case including the conduction.
The Lorentz terms dominate over the baroclinic and inertia terms and balance with the viscous term.
The onset of the explosive reconnection coincides with the enstrophy excitation at $T=180$.
Figure \ref{fig:voz_prof}(b-e) shows the two-dimensional distribution of the three dominant terms and the whole term close to the reconnection site at the onset.
The enstrophy is predominantly excited in the current sheet (red-colored region in (b)).
The second Lorentz term (c), which is approximated as $-(\omega_z j_z/\rho^2) B_y \partial \rho / \partial y$ around the reconnection region, is positive for the density decreasing from the upstream toward the current sheet (for example, $\omega_z>0, j_z<0, B_y>0, \partial \rho / \partial y > 0$ in first quadrant).
The viscous term (d) is negative in the current sheet, but turns to be positive around the outer edge of the sheet due to viscous transfer.
As a whole (e), the enstrophy emerges also outside of the current sheet.
This is an indication of the positive feedback toward the upstream via viscosity.

\begin{figure}[htbp]
 \includegraphics[clip,angle=0,scale=0.45]{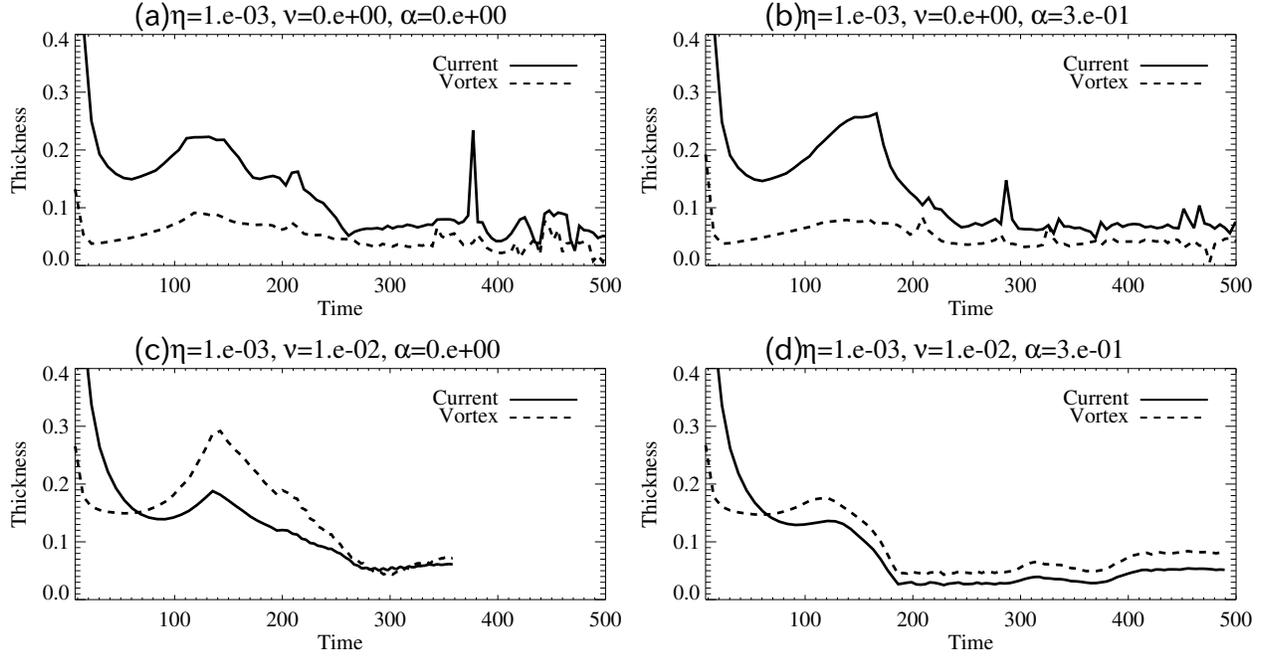}
\caption{Time profile of the thickness of the current (solid lines) and the vortex (dashed lines) layers at the reconnection site in (a) the resistive case, (b) the resistive case including thermal conduction, (c) the visco-resistive case $(Pr_m = 10)$, and (d) the visco-resistive case including thermal conduction $(Pr_m = 10, Pr = 1/30)$.}
\label{fig:thickness_profile}
\end{figure}

Figure \ref{fig:thickness_profile} compares the time profile of the thickness of the out-of-plane current and vortex layers.
They are obtained by fitting the current and vortex distributions across the reconnection site with $\cosh^{-2}(y)$ and $y \cosh^{-2}(y)$ functions.
The ideal MHD condition is violated in the current sheet.
In the resistive case (a), the vortex is restricted within the current sheet.
Since the magnetic field is no longer frozen into the plasma within this vortex scale, the vortex flow does not carry the magnetic flux toward the reconnection site.
{Thermal conduction does not change this situation as is shown in (b).}
On the other hand, the condition $Pr_m > 1$ allows the transfer of the vortex outside of the current sheet as is indicated in Figure \ref{fig:voz_prof}(e).
The vortex can be decoupled from the fluid in the upstream whereas the magnetic field is still coupled with.
Consequently, the vortex is broader than the current in the visco-resistive cases (c,d).
This is a favorable situation for the reconnection because the broad vortex flow enables the transfer of the upstream frozen-in magnetic flux toward the reconnection region.


Viscosity has linear and nonlinear stabilizing effects on the reconnection.
The viscous dissipation leads to plasma heating via $\partial P/\partial t = (\gamma-1) \rho \nu (\vect{\rm S} \cdot \nabla) \vect{u} \simeq (\gamma-1) \rho \nu |\nabla \times \vect{u}|^2$. 
Without convection or diffusion, the heated plasma stagnates at the current sheet and inhibits its thinning.
The thinning speed of the current sheet is slower than that of the vortex in the visco-resistive case without the conduction (Figure \ref{fig:thickness_profile}(c)).
The evolution is saturated when the thickness of the current sheet exceeds the vortex at $T = 280$.
The reconnection rate remains slow in this case.

The simulation indicates that the visco-resistive reconnection can evolve against the inhibition in the presence of thermal conduction.
{Thermal conduction will decrease the temperature around the reconnection region that leads to the compression of the plasma to satisfy the force balance.
It can be accompanied by the upstream magnetic flux provided that the spatial scale of the fluid is broader than that of the magnetic field.}
Figure \ref{fig:thickness_profile}(d) indicates that the thinning speed of the current is as much as that of the vortex and the vortex layer is always broader than the current in the nonlinear stage, persisting the efficient supply of the upstream magnetic flux to the reconnection region.
Both the vortex and current layers are narrower than those in the visco-resistive case without the conduction.
The ratio of the thickness of vortex to current is found to scale as $Pr_m^{1/4}$ within the explored range.
This would be related to the positive correlation of the reconnection rate with viscosity.
The scaling deviates from a simple prediction $Pr_m^{1/2}$ in which the viscous dissipation time balances with the resistive dissipation time.

\begin{figure}
\includegraphics[clip,angle=0,scale=0.4]{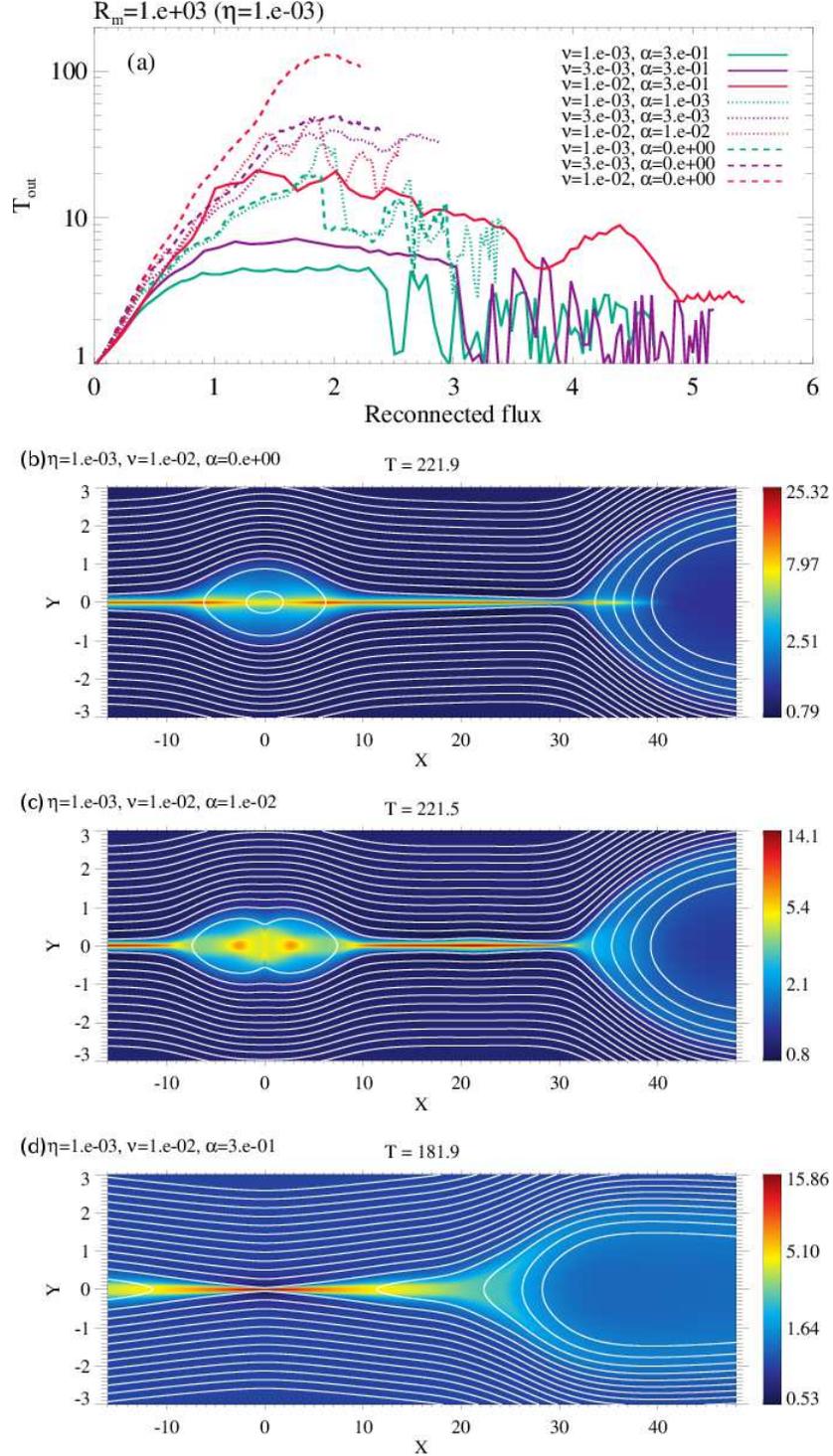}
\caption{(a) Temperature at the reconnection site as a function of the amount of reconnected flux. The various colors represent simulation results with different viscosity. The solid, dotted, and dashed lines denote the results including thermal conduction with $Pr < 1$ and $Pr = 1$, and excluding the conduction $(Pr = \infty)$. (b-d) Two-dimensional temperature distribution at which the reconnected flux is 1.0. The Prandtl numbers are $Pr_m=10$ and $Pr=\infty,1,1/30$.}
\label{fig:temp_profile}
\end{figure}

{
It is reasonable to speculate that $Pr<1$ is a necessary condition to sustain the visco-resistive reconnection since thermal conduction is expected to be fast enough to take away the thermal energy increased by the viscous dissipation.
In order to confirm the effect of thermal conduction, Figure \ref{fig:temp_profile}(a) shows the temperature at the reconnection site as a function of the amount of the reconnected flux.
The peak temperature is roughly proportional to the viscosity.
When $Pr_m=1,3$ (green and magenda), thermal conduction is not effective for the cases $Pr \geq 1$ whereas it successfully decreases the temperature and boosts the reconnection for $Pr < 1$.
When $Pr_m = 10$ (red), on the other hand, thermal conduction succeeds in decreasing the temperature even with $Pr = 1.$
The peak temperature values are 200 for $Pr = \infty$, 40 for $Pr = 1$, and 20 for $Pr = 1/30$.
Figure \ref{fig:temp_profile}(b)-(d) shows the temperature distribution at which the reconnected flux is about the same for these three cases.
High temperature plasma is observed in a quite narrow layer in the case (b), indicating a lack of a way to redistribute the thermal energy.
The temperature distribution in the case (c) is somewhat diffusive.
However, narrow high temperature layer and a large plasmoid remains around $x=10-30$ and $x=0$, and the subsequent evolution is similar to the case (b).
Thus, thermal conduction with $Pr = 1$ is not fast enough to sustain the visco-resistive reconnection compared with the case (d), in which high temperature region becomes broader toward the downstream.
 }

\section{DISCUSSION}\label{sec:discussion}
\begin{figure}[htbp]
 \includegraphics[clip,angle=0,scale=1.0]{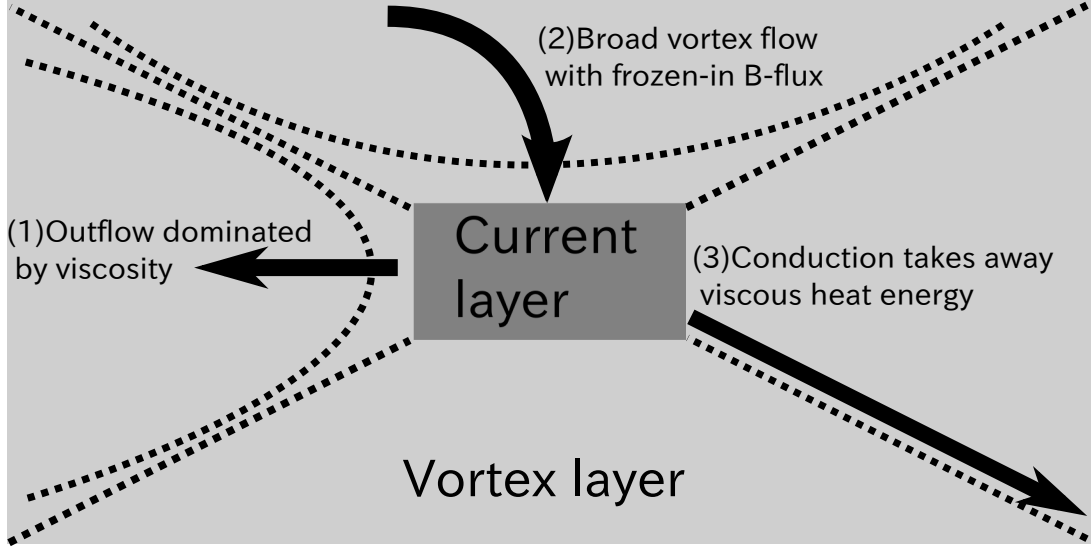}
\caption{Schematic representation of the viscosity-dominated reconnection. The light gray region corresponds to the vortex layer, and the dark gray region is the embedded current layer for $Pr_m > 1$. The dashed lines represent the magnetic field line. (1) Outflow dynamics is dominated by viscosity, leading to the decrease in the plasma density and the vortex excitation. (2) The broad vortex flow can carry the upstream frozen-in magnetic flux toward the reconnection region that boosts the reconnection. (3) Fast thermal conduction $(Pr < 1)$ is required to sustain the boost.}
\label{fig:schematic_rep}
\end{figure}

Based on the visco-resistive MHD simulation coupled with anisotropic thermal conduction, we propose the viscosity-dominated reconnection model in Figure \ref{fig:schematic_rep}.
Viscosity and thermal conduction can be a key to boost the reconnection in the MHD regime.
The reconnection rate is found to increase with viscosity within the explored range provided that thermal conduction is fast enough.
However, the reconnection rate in the present model is still slower than that in kinetic models \citep{2011PhPl...18l2108Z,2015CoPhC.187..137M}.
The Hall-MHD model is thought to be a minimal model for fast reconnection \citep{2001JGR...106.3737B}.
\citealt{1983GeoRL..10..475T} carried out a linear analysis of the resistive tearing instability including the Hall effect, and argued that the thickness of the vortex is broader than the current and the broad vortex flow is expected to enhance the reconnection.
One of differences between the two models is the presence or absence of dissipation.
The Hall effect is purely dispersive mode without dissipation.
It should more efficiently convert magnetic energy (current) into kinetic energy (vortex) than the viscosity-dominated reconnection.

The inflow speed is characterized by the whistler rather than the {\Alfven} speed in the kinetic regime, $u_{\rm in} \sim V_{\rm A,in} d_{\rm i}/L$, where $d_{\rm i}$ is the ion inertia length \citep{1995PhRvL..75.3850B,1998GeoRL..25.3759S,1999GeoRL..26.2163S}.
{If we assume $u_{\rm out} \sim V_{\rm A,in}$ and relate the ratio $V_{\rm A,in}/u_{\rm in}$ to the aspect ratio of the reconnection region, the viscosity-dominated model (eq. (\ref{eq:16})) gives the effective viscosity for the kinetic reconnection as $\nu_{\rm eff} \sim u_{\rm in}L (u_{\rm in} / V_{\rm A, in}) \sim V_{\rm A,in}d_{\rm i} (d_{\rm i}/L)$.
It also gives the effective resistivity as $\eta_{\rm eff} \sim u_{\rm in} \delta \sim V_{\rm A,in}d_{\rm i} (\delta/L)$.}
The effective magnetic Prandtl number $Pr_{m,{\rm eff}}\sim d_{\rm i}/\delta$ may be larger than unity in the kinetic reconnection in which the thickness of the current sheet gets thinner than the ion inertia length (down to the electron inertia length).
{Transition from the slow resistive MHD to the fast Hall-MHD reconnection can be observed when the current sheet thickness falls below the ion inertia length\cite{2010PhRvL.105a5004S}.}

The decreasing density distribution from the upstream toward the current sheet is not a situation observed only in the viscosity-dominated reconnection.
It has been extensively studied that ad hoc localized resistivity triggers fast reconnection in the MHD regime, the so-called Petscheck-type reconnection.
The decreasing density distribution is observed in the Petscheck-type reconnection due to fast magnetosonic rarefaction waves emanated from the reconnection site\citep{2001ApJ...549.1160Y}.
The rarefaction wave drives the upstream plasma toward the reconnection site since $\nabla \cdot \vect{u} \sim -(u_y/\rho) \partial \rho/\partial y > 0$.
The dilatation in the upstream is also seen in the viscosity-dominated reconnection.
Furthermore, thermal conduction facilitates the Petscheck-type reconnection\cite{1999ApJ...513..516C}.
Observation of the solar flare supports these theoretical models \citep{1996ApJ...456..840T}.

Currently, it remains unclear whether the viscosity-dominated reconnection is controlled by diffusion coefficients $(\eta,\nu,\alpha)$ or their ratio $(Pr_m,Pr)$ (or both), because we have fixed the resistivity value. 
Subsequent works will investigate the dependence on resistivity (Lundquist number), which is a key parameter to classify the reconnection dynamics \citep{2011PhPl...18k1207J}.
We may anticipate that the Prandtl numbers control the dynamics to some extent because $Pr_m > 1$ leads to the two-scale structure of the reconnection region and $Pr < 1$ is required to sustain the boost.
This implies that one cannot ignore viscosity and thermal conduction whenever they exceed resistivity, even if their absolute values are small.
The condition $Pr_m \gg 1$ and $Pr \ll 1$ can be expected in actual plasma environments such as the solar atmosphere and the interstellar medium \citep{2015ApJ...801..145T}.
The result indicates the importance of viscosity and heat transfer for the reconnection against the conventional resistive MHD model in which the Prandtl numbers are not explicitly defined.


\begin{acknowledgements}
We thank the anonymous referee for carefully reviewing our manuscript and giving comments that improved the manuscript.
We also thank S. Hirose, Y. Kawamura, and T. Yokoyama for discussions and comments.
The authors are supported by JSPS KAKENHI Grant Numbers JP15K05369 (T. Minoshima), JP15K04756 (T. Miyoshi), and JP26287143, JP15H05816 (S. Imada).
Numerical simulations were in part carried out on Cray XC30 at Center for Computational Astrophysics, National Astronomical Observatory of Japan.
\end{acknowledgements}

%

\clearpage

\end{document}